# Temperature-sensitive spatial distribution of defects in PdSe$_2$ flakes

Xiaowei Liu[1,∥], Yaojia Wang[1,∥], Qiqi Guo[2,∥], Shijun Liang[1,∥], Tao Xu[3], Bo Liu[4], Jiabin Qiao[2], Shengqiang Lai[1], Junwen Zeng[1], Song Hao[1], Chenyi Gu[5], Tianjun Cao[1], Chenyu Wang[1], Yu Wang[1], Chen Pan[1], Guangxu Su[1], Yuefeng Nie[5], Xiangang Wan[1], Litao Sun[3], Zhenlin Wang[1], Lin He[2,*], Bin Cheng[1,6*] and Feng Miao[1,*]

[1]*National Laboratory of Solid State Microstructures, School of Physics, Collaborative Innovation Center of Advanced Microstructures, Nanjing University, Nanjing 210093, China*

[2]*Center for Advanced Quantum Studies, Department of Physics, Beijing Normal University, Beijing, 100875, China*

[3]*SEU-FEI Nano-Pico Center, Key Laboratory of MEMS of Ministry of Education, School of Electronic Science and Engineering, Southeast University, Nanjing 210018, China*

[4]*College of Mechanical and Vehicle Engineering, Hunan University, Changsha, China*

[5]*National Laboratory of Solid State Microstructures, College of Engineering and Applied Sciences, and Collaborative Innovation Center of Advanced Microstructures, Nanjing University, Nanjing 210093, China*

[6]*Institute of Interdisciplinary Physical Sciences, School of Science, Nanjing University of Science and Technology, Nanjing 210094, China*

**Corresponding Authors**

*Email: helin@bnu.edu.cn, *Email: bincheng@njust.edu.cn, *Email: miao@nju.edu.cn

**ABSTRACT: Defect engineering plays an important role in tailoring the electronic transport properties of van der Waals materials. However, it is usually achieved through tuning the type and concentration of defects, rather than dynamically reconfiguring their spatial distribution. Here, we report temperature-sensitive spatial redistribution of defects in PdSe$_2$ thin flakes through scanning tunneling microscopy (STM). We observe that the spatial distribution of Se vacancies in PdSe$_2$ flakes exhibits a strong anisotropic characteristic at 80 K, and that this orientation-dependent feature is weakened when temperature is raised. Moreover, we carry out transport measurements on PdSe$_2$ thin flakes and show that the anisotropic features of carrier mobility and phase coherent length are also sensitive to temperature. Combining with theoretical analysis, we conclude that temperature-driven defect spatial redistribution could interpret the temperature-sensitive electrical transport behaviors in PdSe$_2$ thin flakes. Our work highlights that engineering spatial distribution of defects in the van der Waals materials, which has been overlooked before, may open up a new avenue to tailor the physical properties of materials and explore new device functionalities.**

## I. INTRODUCTION

Defect engineering is a powerful tool for tailoring the electronic transport properties of materials, and plays a key role in the silicon-based devices. A typical example of defect engineering is the implantation of donors and acceptors in the bulk silicon to form p–n junctions [1], which lays the essential foundation for the semiconductor devices and applications [2]. In the last decade, defect engineering on thin flakes of van der Waals (vdW) materials has sparked much interests since the reduced dimensionality can enhance the impact of defects on the physical and chemical properties of materials [3]. Much efforts have been made to generate and manipulate defects in thin flakes of vdW materials, such as chemical doping [4], electron irradiation [5], plasma treatment [6] and thermal annealing [7], etc. However, it should be noted that these strategies for defect engineering in vdW materials only concern the type and concentration of defects [3], while overlook other defect degrees of freedom. As an alternative approach, dynamically reconfiguring spatial distribution of lattice defects, which is tightly related to lattice symmetry of materials [8,9], can be utilized for engineering physical properties and customizing device functionalities in vdW materials.

As a representative of layered noble-transition metal dichalcogenide semiconductors, PdSe$_2$ has recently received extensive attention due to pressure tunable conduction and the field emission [10,11], and defect-mediated exotic phenomena such as interlayer fusion [12], phase transition [13-15], photo-controllable adsorption and desorption [16],etc. Different from symmetric lattice structures of thin flake transition metal dichalcogenides among the vdW materials, PdSe$_2$ thin flake has a relatively low crystal symmetry (point group: D$_{2h}$) [17-19], which results from the orthorhombic lattice structure with one palladium atom coordinated with four selenium atoms, as shown in Fig. 1(a). Prior works have identified Se vacancies (V$_{Se}$) as typical defects in PdSe$_2$ and demonstrated electrical field driven V$_{Se}$ migration [20]. Notably, the theoretical calculations show that the diffusion barrier of defect is as low as 0.03 eV in PdSe$_2$ [20] and is comparable to the thermal energy at room temperature. With the low-symmetry lattice structures, temperature variation induced migration of Se vacancies (V$_{Se}$) could significantly modulate the spatial characteristics of defect distribution in PdSe$_2$ and provide an opportunity to tailor the physical properties.

In this work, we investigate the temperature-dependent spatial distribution of V$_{Se}$ in PdSe$_2$ thin flakes by employing scanning tunneling microscopy (STM). We show that a strong anisotropic characteristic of V$_{Se}$ distribution appears at 80 K, and gradually disappears with increasing temperature. Furthermore, we demonstrate that such orientation dependence of V$_{Se}$ distribution gives rise to temperature-sensitive anisotropic behaviors of carrier mobility and phase coherent length along different crystallographic orientations of PdSe$_2$ flakes. Our work may highlight that tuning the spatial distribution of defects can play an important role in engineering physical properties of vdW materials.

## II. EXPERIMENTAL METHODS

**Synthesis of PdSe$_2$ single crystal.** Single crystals of PdSe$_2$ were synthesized by self-flux method [21]. The selenium powder (99.999%) and palladium powder (99.95%) were mixed with an atomic ratio of 2:1, and compressed into tablets and transferred to a quartz tube sealed under $10^{-5}$ mbar. After that, the sealed quartz was placed in a muffle furnace, which was then slowly heated to 850℃ and held for 70 h to make the chemical reaction fully proceeded. After the muffle furnace was cooled down to room temperature, we took out the poly-crystalline samples of PdSe$_2$, and mixed them with Se powder with a mass ratio of 1:4, and sealed in another evacuated quartz tube sealed under $10^{-5}$ mbar and transfer the tube into a muffle furnace. Then we again slowly heat the furnace to 850℃ and held for 70 h. After the furnace was cooled down to room temperature again, single crystals of PdSe$_2$ obtained by cleaving the excess Se ingot.

**Characterizations.** The chemical composition of PdSe$_2$ samples were measured by Hitachi S-3400 N II SEM EDXS. The crystal structures were characterized by the Bruker D8 Discover X-ray diffractometer with Cu K$_α$ radiation (λ = 1.5406 Å) operating in Bragg-Brentano mode. Raman spectrum was measured under 514 nm laser through Renishaw inVia Raman microscope H54304. High-resolution TEM (Titan 80–300) and SAED were performed at the accelerating voltage of 200 kV.

**Device Fabrications and Measurements.** PdSe$_2$ thin flakes were obtained by using a standard mechanical exfoliation method, and transferred to SiO$_2$ (300nm)/Si substrate. The metal electrodes (5 nm Ti/45 nm Au) were fabricated using electron-beam lithography followed by electron-beam evaporation. The devices were measured in an Oxford cryostat with a magnetic field of up to 4 T and based temperature of about 1.6

K. Gate-dependent four terminal conductivity measurements were performed by using a Keithley 2636 and 2400 digital source meter. Magneto-transport measurements were performed by using a low-frequency Lock-in amplifier.

**STM/STS measurements.** $PdSe_2$ samples with freshly cleaved surfaces were rapidly transferred into ultrahigh vacuum chamber (~$10^{-8}$ Torr) after mechanical exfoliation. STM/STS measurements were performed in the ultrahigh vacuum chamber (~$10^{-11}$ Torr) with constant-current scanning mode through a scanning probe microscope (USM-1400) from UNISOKU. Lateral dimensions in the STM images were calibrated using standard graphene lattice, Si (111)-(7×7) lattice and Ag (111) surface. The STM tip is grounded during measurement, and the bias voltage $V_{ds}$ was applied on the $PdSe_2$ samples in the STM measurements.

## III. RESULTS AND DISCUSSION

We used the method of self-flux to synthesize $PdSe_2$ single crystals [Fig. 1(b)] and adopted standard mechanical exfoliation to obtain $PdSe_2$ thin flakes for experimental study. The orthorhombic $PdSe_2$ phase was confirmed by the X-ray diffraction (XRD) pattern (PDF card *No.*11-0453), as shown in Fig. 1(c). The chemical composition was determined by energy dispersive X-ray spectroscopy (EDXs), as shown in Fig. 1(d). The atomic ratio between Pd and Se is about 1:1.88, indicating the existence of abundant Se vacancies in the as-grown $PdSe_2$ samples. We determined the in-plane crystalline orientation of the exfoliated $PdSe_2$ thin flakes by using angle-resolved polarized Raman spectroscopy (ARPRS) and high-resolution transmission electron microscopy (HRTEM) [22], given that the physical properties of low-symmetry materials usually depend on the crystalline orientation. To carry out the ARPRS measurement on $PdSe_2$ samples, we kept the polarization of Raman scattered light parallel to that of the incident laser and rotated the sample. The intensity of $A_g^1$-$B_{1g}^1$ mode exhibits strong angular dependence [see Fig. S1 [23]] and an obvious periodic variation period of 180° [inset of Fig. 1(e)]. Combining the ARPRS results with TEM image [Fig. 1(f)] based on the same sample, we reveal that the strength of $A_g^1$-$B_{1g}^1$ mode reaches its maximum (minimum) when the polarization of incident laser is parallel to *a*-axis (*b*-axis) of PdSe$_2$ thin flakes. This justifies that ARPRS measurements can be used to determine the crystalline orientations of other PdSe$_2$ samples used in the experiments.

We employed high-resolution scanning tunneling microscopy to characterize the atomic structure of the cleaved surfaces of PdSe$_2$ flakes, with the atomically resolved STM image shown in Fig. 2(a). The white spots that consist the square cells [see lower right in Fig. 2(a)] represent the Se atoms, which are consistent with Se lattice of PdSe$_2$ (001) surface [see upper right in Fig. 2(a)]. Se vacancy defects are identified to be the yellow bright dots surrounded by white circles as shown in Fig. 2(a) [see Fig. S2 [23]]. The existence of Se vacancy defects confirmed by the STM measurements is consistent with our EDX results and previous work [12,20]. Notably, most of the $V_{Se}$ are distributed discretely, while a small fraction of $V_{Se}$ form unique chain-like structures along b-axis at low temperature [see Fig. S3 [23]]. We characterized the spatial distribution of the $V_{Se}$ by counting the numbers of $V_{Se}$ along different axes. Specifically, we drew evenly spaced line grids along *a*- and *b*-axis in large-scale STM images visualizing $V_{Se}$, and counted the numbers of $V_{Se}$ along each grid line [see Fig. S4 [23]]. Statistical analysis was performed on tens of such STM images taken from different regions of PdSe$_2$ surface, with the effective area larger than 10,000 nm$^2$. Representative STM images and the statistical histograms of $V_{Se}$ distribution at 80 K and 110 K are shown in Fig. 2(c) and Fig. 2(d), respectively. By fitting the statistical histograms at 80 K with Poisson distribution function, we estimated the average number of $V_{Se}$ in a grid line along *a*- and *b*-axis to be 1.7±0.2 and 3.7±0.3, respectively. These distinct expected values suggest a strong anisotropic characteristic of $V_{Se}$ distribution along different axes, in consistence with the results shown in Fig. 2(b). We also analyzed the statistical histograms at 110 K and obtained the average numbers of $V_{Se}$ along *a*- and *b*-axis, which are 3.1±0.2 and 3.9±0.6, respectively. Compared to those values at 80 K, the difference between the average numbers of $V_{Se}$ along *a*- and *b*-axis at 110 K are largely diminished, indicating that increasing temperature would weaken the anisotropic characteristic of $V_{Se}$ distribution. This temperature-sensitive orientation dependence of $V_{Se}$ distribution could be understood by the competition between energy and entropy. When temperature is low, a lower energy is required to align $V_{Se}$ along *b*-axis compared to that along *a*-axis, which is confirmed by the first-principles calculations [see Calculations about the total energy of PdSe$_2$ with Se vacancies [23]]. As temperature increases, the increased entropy intends to distribute $V_{Se}$ isotropically along *b*- and *a*-axis. This dynamical modulation of $V_{Se}$ distribution controlled by temperature varying is enabled by the extremely low defect diffusion barrier in PdSe$_2$ (~0.03 eV), which is much lower than that reported in other vdW materials such as MoS$_2$ (2.3 eV for S vacancies [24]) and BN (2.6 eV for B vacancies [25]).

It is well known that defect-induced scattering effects on charge carriers could lead to profound modulation of electrical transport behaviors in thin flakes of vdW materials [1,3]. Therefore, the tuning of defect spatial distribution in PdSe$_2$ could be used as a new knob for engineering the electrical transport properties. To that end, we fabricated PdSe$_2$ field-effect device (10~15 nm) to investigate the temperature dependence of electrical transport properties. Figure 3(a) and inset of Figure 3(b) respectively show the schematic and corresponding optical image of the PdSe$_2$ device along both *a*- and *b*-axis. The transfer curve of the device displays an ambipolar carrier transport behavior with a lightly n-type doping at room temperature, which might result from the Se defects [see Fig. S5 [23]]. We measured gate-dependent four terminal conductivity along *a*- and *b*-axis at different temperatures, which were respectively marked as $\sigma_a$ and $\sigma_b$, and presented the results in Fig. 3(b) and 3(c). At low temperature, $\sigma_a$ exhibits an oscillating feature, which might be a signature of defects induced density of states [26-28]. Figure 3(d) shows the conductivities along *a*- and *b*-axis versus temperature with the gate voltage kept at 70 V. As the temperature decreases, the difference in conductivity between *a*-axis and *b*-axis gradually increases. We extract the ratio of $\sigma_a$ to $\sigma_b$ to characterize the anisotropic properties of electrical transport in PdSe$_2$ thin flakes. The ratio varies from 6.2 to 9.9 at 1.6 K for different gate voltages and is drastically reduced at high temperature, *i.e.* 1.5~2.4 at 250 K [see Fig. S6 [23]]. Based on the measured four-terminal conductivity, we extracted field-effect carrier mobilities along two different axes (marked as $\mu_a$ and $\mu_b$, respectively) and observed similar temperature dependence as well: $\mu_a/\mu_b$ is about 7.2 at 1.6 K, and drops to ~1.1 at 250 K. Here, the field-effect mobilities $\mu_a$ and $\mu_b$ were extracted according to $\mu = \frac{d}{\varepsilon_0 \varepsilon_r} \times \frac{dI_{ds}}{dV_g} \times \frac{1}{V_{ds}} \times \frac{L}{W}$, where d denotes the thickness of SiO$_2$ layer, $\varepsilon_0$ and $\varepsilon_r$ are the vacuum permittivity and the SiO$_2$ dielectric constant ($\varepsilon_r$ = 3.9), respectively. Notably, we exclude the influence of sample aspect ratio by considering the factor of *L/W* when calculating the mobility. In addition to the scattering effect from anisotropically distributed defects, the intrinsic anisotropy of carrier effective mass caused by low crystalline symmetry may also contribute to the anisotropic mobility. However, this possibility can be ruled out since the effective masses (*i.e.* 0.30$m_e$ and 0.36$m_e$) along *a*- and *b*-axis in PdSe$_2$ flakes are very close to each other according to the first-principles calculations [see DFT band structures calculations [23]]. Besides, the weak orientation dependence of the effective mass also fails to interpret the temperature-sensitive characteristics of $\mu_a/\mu_b$. Therefore, it is safe to attribute such unusually varying feature of carrier mobility with temperature to the temperature-driven spatial redistribution of vacancies, which act as moveable scattering centers to engineer the carrier transport. Notably, the carrier mobility of PdSe$_2$ reflect the transport properties of the sample bulk, so the strong anisotropy of carrier mobility indicates that the anisotropic distribution of V$_{Se}$ is definitely a bulk property. Moreover, we can also estimate the order/disorder transition temperature of defect distribution to be around 200 K, at which the mobility along *a*- and *b*-axis become equal.

Besides carrier mobility, temperature-driven spatial redistribution of vacancies also allows for modulation of the phase coherence length. Since electrons experience random scattering of vacancies and lead to many different random paths, those self-crossing paths would generate quantum interference at low temperature. As a result, the temperature-driven spatial redistribution of vacancies would modulate the electron phase coherence length along different crystallographic orientations of PdSe$_2$ material. To gain the insight into the modulation of phase coherence length by temperature-sensitive spatial distribution of vacancies, we performed magneto-transport measurements along *a*- and *b*-axis at different temperatures, with results shown in Fig. 4(a) and 4(b) respectively. The perpendicular magnetic field results in positive magneto-conductivities ($\Delta\sigma(B) = (L/W)(1/R(B) - 1/R(0))$) along both *a*- and *b*-axis at several different gate voltages at 1.6 K, where $R(B)$ is the magnetic field dependent resistance, and $L$ and $W$ are the channel length and width, respectively. Such feature is consistent with weak localization phenomenon [29-31], which manifests itself in the conductivity correction due to quantum phase coherence effect. The phase coherence length $L_\varphi$ can be extracted by fitting the Hikami–Larkin–Nagaoka (HLN) equation [32,33]:

$$\Delta\sigma = \frac{e^2}{\pi h}\left[\Psi\left(\frac{1}{2} + \frac{B_\varphi}{B}\right) - \ln\left(\frac{B_\varphi}{B}\right)\right] \quad (2)$$

Here, $h$ is the Planck's constant, $e$ corresponds to elemental charge, $\Psi$ refers to digamma function, and $B_\varphi = h/(8\pi e L_\varphi^2)$ represents the characteristic field of phase coherence. Figure 4(c) shows the fitted $L_\varphi$ dependence of gate voltage along *a*- and *b*-axis. These two $L_\varphi$ curves exhibit increasing trends along with the increase of the gate voltage, indicating that the scatterings between carriers and vacancies become weak due to high carrier density enhanced screening effects. At the temperature of 1.6 K, $L_\varphi$ along *a*-axis is 2.4 ~ 2.6 times larger than that along *b*-axis for various gate voltages. When temperature is raised to 25 K, the ratio of $L_\varphi$ along *a*- and *b*-axis is reduced down to ~1.9, as shown in Fig. 4(d) [details see Fig. S8 [23]]. Since $L_\varphi$ is inversely related to the scattering strength, the suppression of the anisotropy of $L_\varphi$ with increasing temperature indicates that more scattering centers are available along *b*-axis to break electron phase coherence than that along *a*-axis. As we mentioned above, the movable vacancies serve as the scattering centers in PdSe$_2$ thin flakes. Therefore, such picture is consistent with temperature-sensitive orientation dependence of carrier mobility as well as the temperature-driving spatial redistribution of Se vacancies observed in STM images.

## IV. CONCLUSION

In summary, we observed anisotropic distribution pattern of Se vacancies in PdSe$_2$ thin flakes at low temperature, which is gradually weakened as temperature increases. Moreover, we revealed that the temperature-sensitive orientation dependence of carrier mobility and phase coherent length can be attributed to temperature-driven rearrangement of the defect distribution in PdSe$_2$ thin flakes. Our work has proposed a new type of defect engineering approach beyond the conventional strategy of tuning the defect type or concentration, and may offer a different avenue for creating novel device functionalities.

## V. OUTLOOK

Defect engineering is a powerful tool for tuning the electronic transport properties in vdW materials. Methods reported so far mainly rely on the *ex-situ* engineering of defect type and concentration, hindering the realization of new types of device functionalities associated with defect engineering. Therefore, it is highly desirable to develop alternative strategies beyond the engineering of the defect type and concentration. This work demonstrates that low crystalline symmetry and anisotropic diffusion barrier of Se vacancies in PdSe$_2$ enable the engineering of defect redistribution. The spatial characteristics of temperature-driven defect redistribution are confirmed to be strongly related to the transport properties such as anisotropic carrier mobility and phase coherent length. *In-situ* modulation of defect distribution in the low-crystalline symmetry vdW materials allows for achieving distinct conducting states and designing novel-concept devices, which would be next target in the following research. Another promising research direction is to explore other tailoring approaches to replace the temperature-based strategy for the engineering of defect distribution, either in PdSe$_2$ or other vdW materials.

**Acknowledgements**

We would like to thank the fruitful discussion with Dr. Liangbo Liang from Oak Ridge National Laboratory. This work was supported by National Natural Science Foundation of China (61625402, 62034004, 61921005, 61974176, 12074176), and the Collaborative Innovation Center of Advanced Microstructures and Natural Science Foundation of Jiangsu Province (BK20190276, BK20180330), Fundamental Research Funds for the Central Universities (020414380084).

**Financial Interest Statements**

The authors declare no competing financial interest.

**AUTHOR INFORMATION**

**Corresponding Authors**

*Email: helin@bnu.edu.cn

*Email: bincheng@njust.edu.cn

*Email: miao@nju.edu.cn

**Author Contributions**

‖Xiaowei Liu, Yaojia Wang, Qiqi Guo, Shi-Jun Liang, contributed equally to this work. X.L., S.J.L., B.C. and M.F. designed the project. S.J.L, B.C. and M.F. supervised the whole project. X.L. synthesized the samples. Q.G., B.Q. and L.H. carried out STM and analyzed the corresponding data. T.X. and L.S. carried out TEM characterization. S.H. and X.L. carried out EDX. C.G. and Y.N. carried out XRD and analyzed the corresponding data. G.S, X.L. and Z.W. carried out Raman characterization. X.L., C.W., Y.W., and C.P. fabricated devices. X.L. and Y.W. performed electrical transport measurements, and X.L., Y.W., J.Z, S.J.L. and B.C. analyzed the corresponding experimental data. S.L., B.L and X.W. carried out first-principle calculations. X.L., B.C., S.L.J., and F.M. co-wrote the manuscript and all authors contributed to

discussions of the results.

# Figure 1

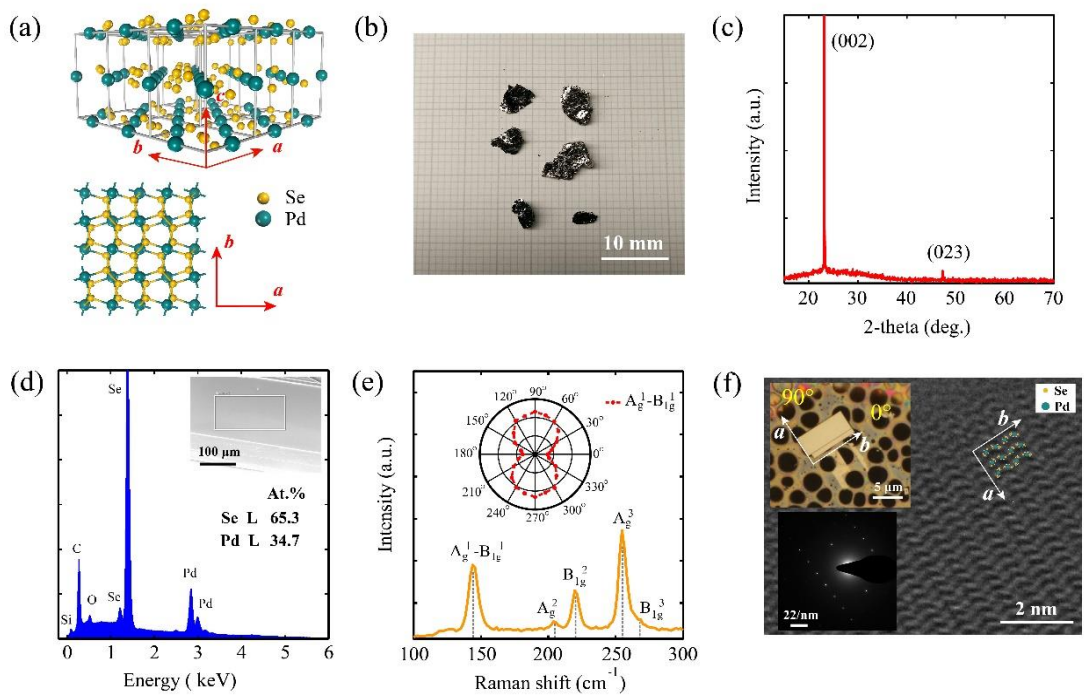

**FIG. 1.** (a) Crystallographic structures of PdSe$_2$. Top panel: Schematics of three-dimensional PdSe$_2$ lattice structure. Bottom panel: the top view. (b) The photograph of PdSe$_2$ single crystals grown by self-flux method. (c) X-ray diffraction (XRD) pattern. (d) Energy dispersive X-ray spectroscopy (EDXs). Inset: The SEM image of one typical PdSe$_2$ single crystal. (e) The Raman spectra of PdSe$_2$ flakes. Inset: the fitted peak intensities of $A_g^1$-$B_{1g}^1$ mode vs sample rotation angle θ under parallel-polarization configuration. (f) High-resolution TEM image. Inset: the optical microscopy image of one PdSe$_2$ flake on the carbon-coated copper mesh (upper left). Selected-area electron diffraction (SAED) pattern of the PdSe$_2$ flake (bottom left).

# Figure 2

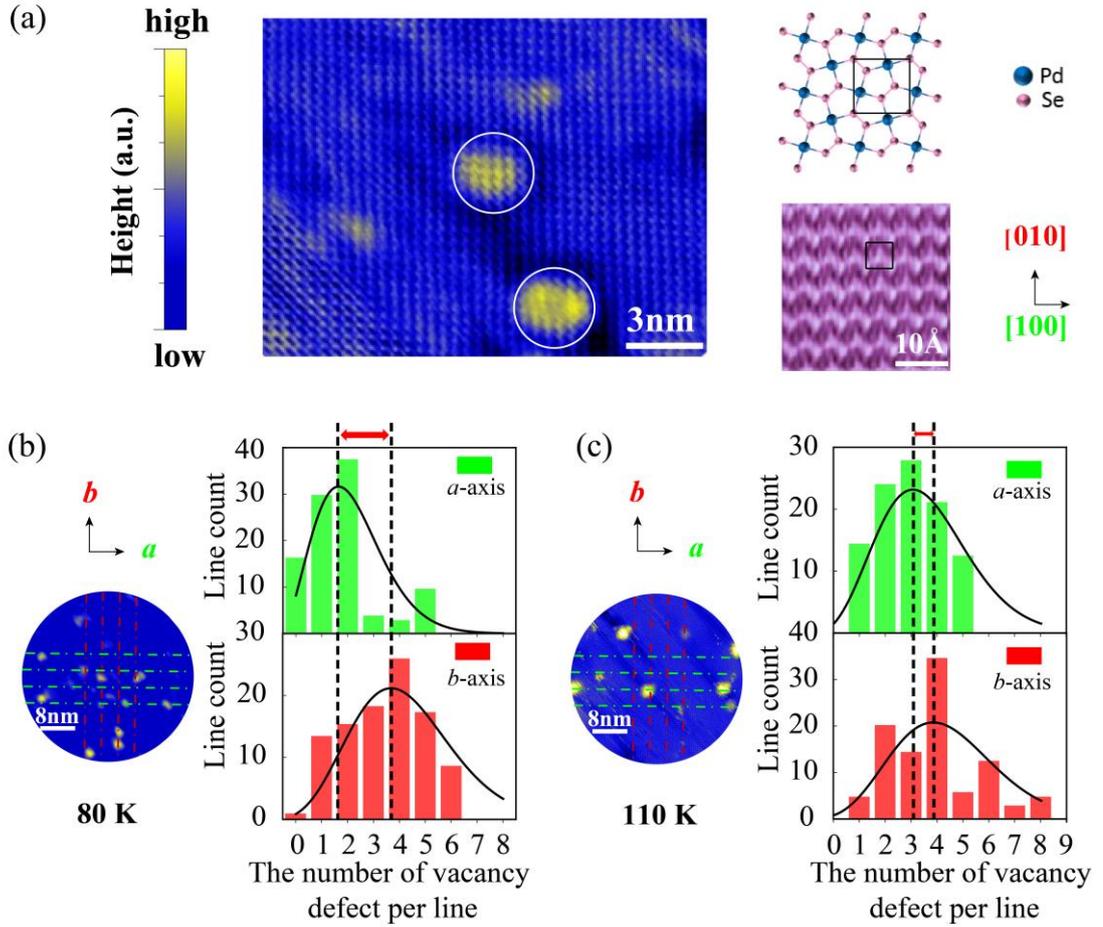

FIG. 2. Characterization of Se-vacancy defects by STM tip. (a) A STM topographic image of PdSe$_2$, where Se vacancy (V$_{Se}$) defects are marked with white circles ($V_s$ = -190 mV, $I$ = 300 pA). Upper right: sketch of the crystallographic structure of PdSe$_2$; Lower right: atomically resolved STM image of PdSe$_2$. (b) – (c) STM topographic images and the statistical histogram of V$_{Se}$ distribution along $a$- and $b$-axis at 80 K and 110 K, respectively (c: $V_s$ = -800 mV, $I$ = 150 pA; d: $V_s$ = -500 mV, $I$ = 150 pA).

# Figure 3

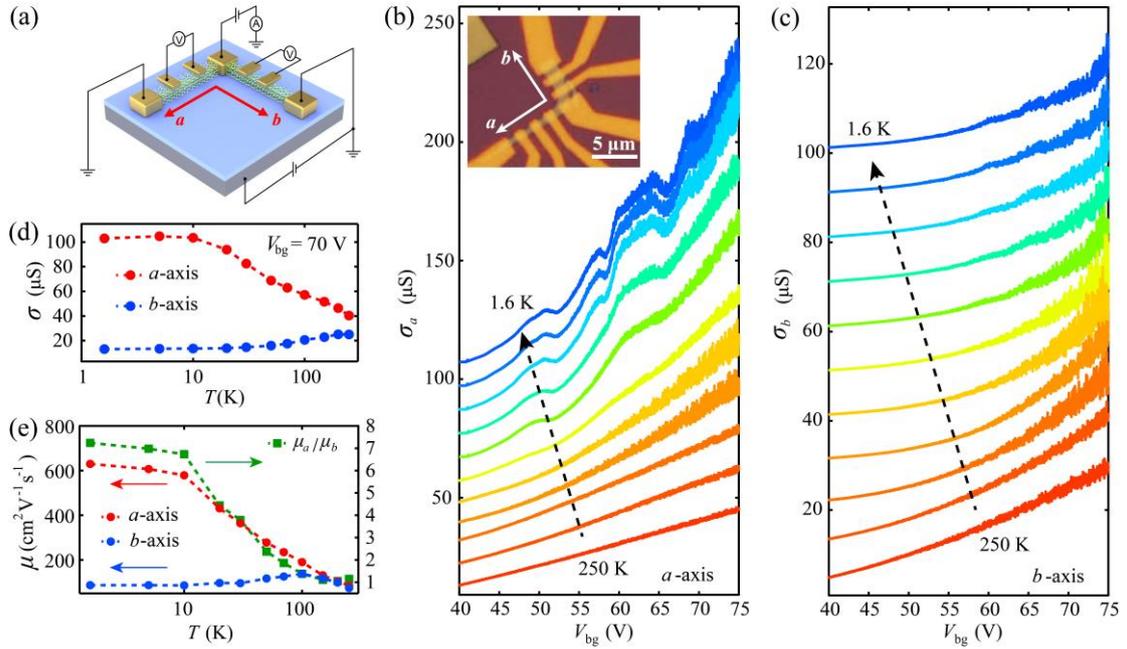

**FIG. 3.** (a) Schematic geometry of PdSe$_2$ device and the transport measurement setup. (b)-(c) Gate dependent conductivity σ from 250 K to 1.6 K along a- and b-axes, respectively. Inset: optical image of PdSe$_2$ device. For clear distinction, the conductivity value of the curve between adjacent temperatures is offset by 10 μS from 250 K to 1.6 K. (d) The conductivities along *a*- and *b*-axes versus temperature, with the gate voltage at 70 V. (e) Mobility along *a*- and *b*-axis, and their ratio versus temperature at a given gate voltage 70 V.

# Figure 4

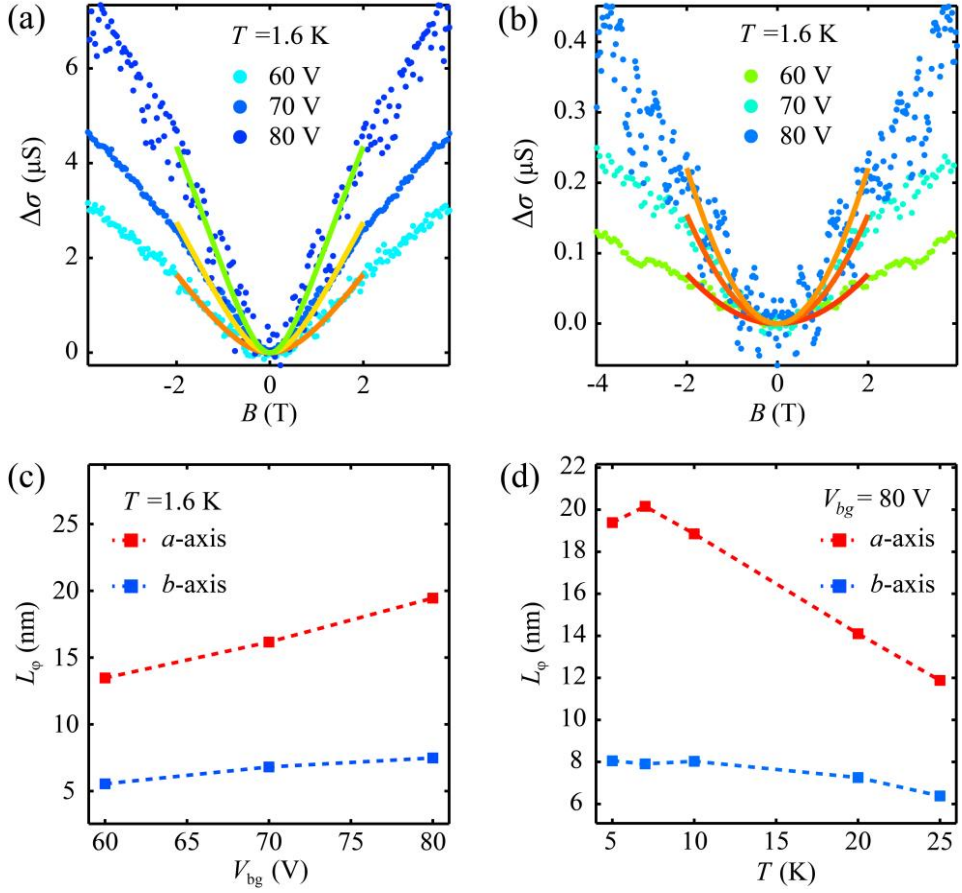

**FIG. 4. Anisotropic quantum interference effects in PdSe$_2$. (a) - (b) The gate dependent magneto-conductivity at 1.6 K with current along *a*- and *b*-axis, respectively. The solid lines are the fitting curves of the magneto-conductivity data with HLN model. (c) The fitted phase coherence lengths $L_\varphi$ vs $V_{bg}$ at 1.6 K. (d) The fitted $L_\varphi$ vs *T* at $V_{bg}$ = 80V.**

# Supplemental Materials

# Temperature-sensitive spatial distribution of defects in PdSe$_2$ flakes

Xiaowei Liu[1,∥], Yaojia Wang[1,∥], Qiqi Guo[2,∥], Shijun Liang[1,∥], Tao Xu[3], Bo Liu[4], Jiabin Qiao[2], Shengqiang Lai[1], Junwen Zeng[1], Song Hao[1], Chenyi Gu[5], Tianjun Cao[1], Chenyu Wang[1], Yu Wang[1], Chen Pan[1], Guangxu Su[1], Yuefeng Nie[5], Xiangang Wan[1], Litao Sun[3], Zhenlin Wang[1], Lin He[2,*], Bin Cheng[1,6*] and Feng Miao[1,*]

[1]*National Laboratory of Solid State Microstructures, School of Physics, Collaborative Innovation Center of Advanced Microstructures, Nanjing University, Nanjing 210093, China*

[2]*Center for Advanced Quantum Studies, Department of Physics, Beijing Normal University, Beijing, 100875, China*

[3]*SEU-FEI Nano-Pico Center, Key Laboratory of MEMS of Ministry of Education, School of Electronic Science and Engineering, Southeast University, Nanjing 210018, China*

[4]*College of Mechanical and Vehicle Engineering, Hunan University, Changsha, China*

[5]*National Laboratory of Solid State Microstructures, College of Engineering and Applied Sciences, and Collaborative Innovation Center of Advanced Microstructures, Nanjing University, Nanjing 210093, China*

[6]*Institute of Interdisciplinary Physical Sciences, School of Science, Nanjing University of Science and Technology, Nanjing 210094, China*

**Corresponding Authors**

*Email: helin@bnu.edu.cn, *Email: bincheng@njust.edu.cn, *Email: miao@nju.edu.cn

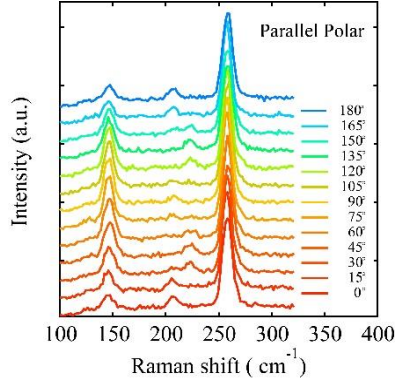

**Fig. S1.** The angle-resolved polarized Raman spectroscopy (ARPRS) of PdSe$_2$ shows that the peak intensity at ~145 cm$^{-1}$ is strongly dependent on the angle between the direction of 0°(marked by the white arrows in Figure 1f inset) and incident light polarization direction under parallel configuration (the incident polarization vector ($e_i$) paralleling to the scattered polarization cvector ($e_s$).).

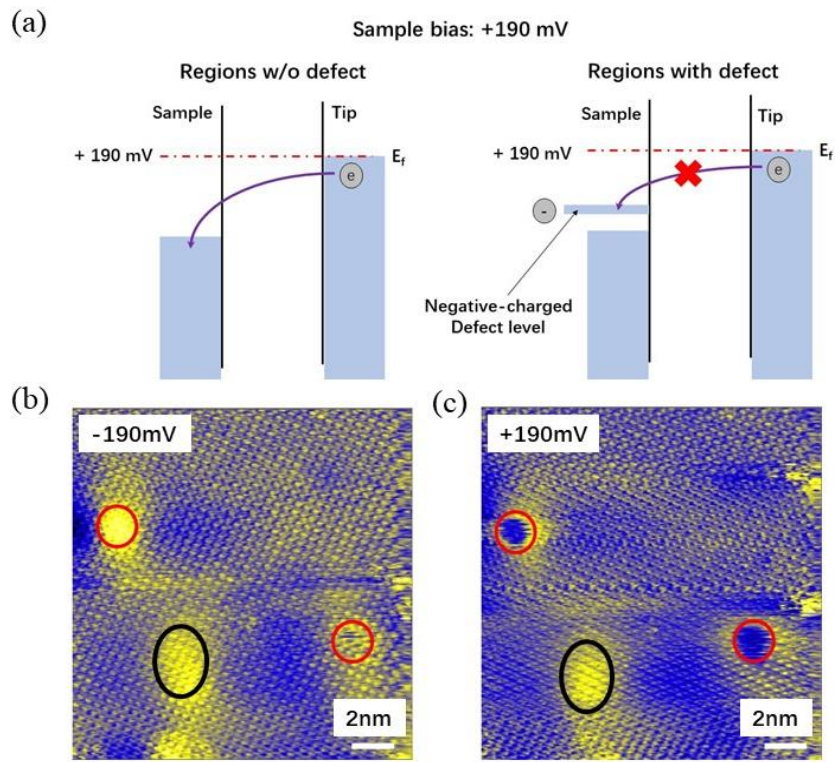

**Fig. S2.** (a) the Schematic diagram of the principle of STM characterization of defects. When the sample has positive bias voltage, as shown in Fig. S2 (a), the tunneling carriers are electrons. As the tip is on top of the defects, the tunneling of the electrons to the negative-charged defects is suppressed due to the electrostatic repulsion

interactions. Therefore, the STM tip should be lowered to keep the current constant in the measurement with constant current mode, leading to a smaller height of the tip and making the defects appear as dark dots, as shown in Fig. S2 (c). On the contrary, when a negative bias is applied to the sample, the charge carriers are holes, which has attraction interactions with the negative defects. Therefore, the STM tip will be lifted to keep the tunneling current constant, making the defects appear as a bright dots, as shown in Fig. S2 (b). Notably, by changing the sign of the bias voltage, we can distinguish the surface topography from the charge defect distribution, just as indicated in the Figure S2 (b) and (c): the red circles represent the defects, whose brightness could be tuned by changing the bias voltage, and the black circle represents the surface topography whose brightness keeps unchanged when bias voltage is tuned from negative to positive. (b) and (c) Morphology of Se vacancy defects using opposite bias voltages in the STM measurements. (b) A representative STM image taken at a negative bias ($V_s = -190$ mV, $I = 200$ pA) where Se vacancy defects are shown as the bright dots. (c) STM image of the same position at a positive bias ($V_s = 190$ mV, $I = 200$ pA), the defects appear as dark spots.

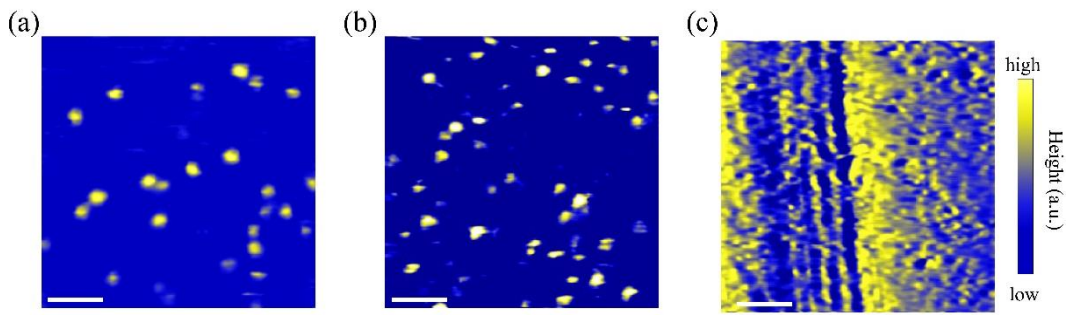

**Fig. S3. Typical STM images** displaying (a) low, (b) intermediate and (c) high concentration of $V_{Se}$ defects, respectively. The scans were performed with $V_s = -550$ mV and $I = 150$ pA. Scale bar in (a)-(c), 10 nm, $T = 80$ K. Notably, high concentration of $V_{Se}$ defects results in the formation of unique chain-like structures along b-axis. Several possible mechanisms such as vacancy-vacancy interaction, asymmetric average energy potential and release of stress could lead to the formation of V chains in PdSe$_2$.

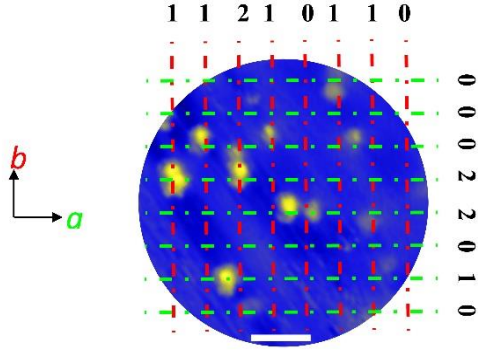

**Fig. S4. Statistical diagram of the number of Se-vacancy defects along different axes.** The green and red lines are parallel to *a* and *b* axis, respectively. Using evenly spaced line grids, the number of $V_{Se}$ defects located along each single line parallel to *a* or *b* axis is counted and labelled. In this way, we obtain the histogram of the number of defects spreading along a single *a* or *b* axis line at two temperatures (80 K and 110 K). The scans were performed with $V_s$ = -550 mV and $I$ = 150 pA. Scale bar, 10 nm.

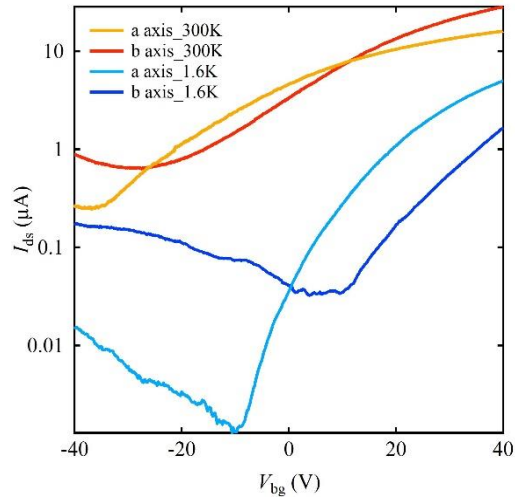

**Fig. S5. Room-temperature and low temperature transfer characteristic near $V_{bg}$ = 0 for the FET with 1V applied bias voltage $V_{ds}$ along a- and b-axis.** The transfer curve of the device displays an ambipolar carrier transport behavior with a lightly n-type doping, which might result from the Se defects. As the temperature decreases, the thermal excitation is partially suppressed, which increases the device on-off ratio.

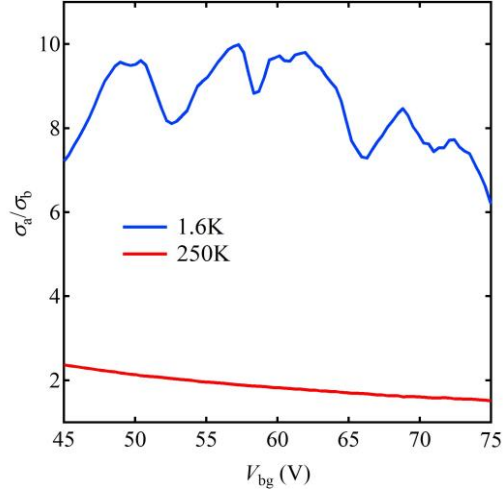

**Fig. S6. $\sigma_a/\sigma_b$ vs $V_{bg}$ at 1.6 K and 250 K** The ratio varies from 6.2 to 9.9 at 1.6 K for different gate voltages and is drastically reduced at high temperature, *i.e.* 1.5~2.4 at 250 K.

**DFT Band Structures Calculations.** PdSe$_2$ band structures were calculated by the density-functional theory using the VASP package.[1] The exchange-correlation and vdW interactions were taken into account by adopting the Perdew-Burke-Ernzerhof (PBE) functional [2] and optPBE-vdW functional,[3] respectively. The meshes of k-point for PdSe$_2$ monolayer and bulk are 9×9×1 and 9×9×9, respectively, with a cut-off energy of 300 eV. The lattice constants were optimized, obtaining a=5.85 Å, b=5.99 Å, c=7.99 Å, which agree with experimental values (a=5.741 Å, b=5.866 Å, c=7.691 Å).[4] With including the Spin-Orbit Coupling (SOC), the band gaps of PdSe$_2$ monolayer and bulk were calculated to be 1.4 eV and 0.09 eV, respectively. The effective electron mass for bulk PdSe$_2$ along *a*-axis and *b*-axis are ~0.30$m_e$ and ~0.36$m_e$, respectively, as obtained by using the formula $m = \hbar^2 (\partial^2 E / \partial k^2)^{-1}$.

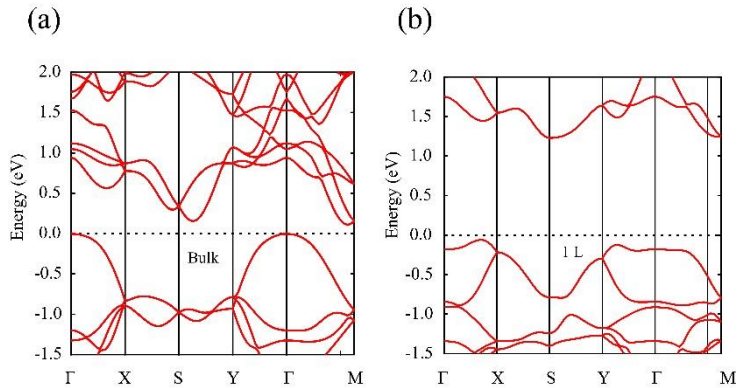

**Fig. S7. The calculated electronic band structures of (a) bulk and (b) monolayer**

**PdSe$_2$.**

**Calculations about the total energy of PdSe$_2$ with Se vacancies**

First principles calculations were performed with the density functional theory (DFT) formalism which is implemented in the plane-wave based Vienna *ab-initio* simulation package (VASP)[5], using the projector augmented wave (PAW) method[6]. The exchange-correlation is included by adopting the Perdew-Burke-Ernzerhof (PBE) version of the generalized gradient approximation (GGA) [2]. The interlayer van der Waals interaction is considered using the optB88-vdW functional. A plane-wave cutoff energy of 500 eV and an energy convergence criterion of $10^{-4}$ eV are adopted in the calculation.

Two supercells with the indexes 2x3x1 and 3x2x1 are constructed. In both supercells, a Se vacancy is created by removing one Se atom. The Se atoms removed in both supercell have the same coordinate with respect to the origin. After duplication along the xyz directions, the defective 2x3x1 supercell results in a much higher vacancy concentration along *a axis* and lower concentration along *b axis*. Both supercells are fully relaxed followed with geometrical optimization until the Hellmann-Feynman forces on each atom are less than 0.01 eV/Å. Calculations are also performed for the perfect 2x3x1 and 3x2x1 supercells for comparison. Calculation results show that after the geometry optimization, the total energy of the perfect 2x3x1 and 3x2x1 supercells is -161.51 eV and -161.48 eV, respectively, while the corresponding total energy for the defective supercells is -158.34eV and -158.93 eV. This indicates that for the defective 3x2x1 supercell structure with a higher vacancy concentration along the *b axis*, the total energy is 0.6 eV lower. Hence, vacancies tend to align along the *b axis*, which is in consistency with the experimental observation.

For perfect 2x3x1 and 3x2x1 supercell structures, the lattice constants after geometry optimization are a=5.85 Å, b= 5.94 Å and c= 7.68 Å. For the defective 2x3x1 supercell, the lattice constant after geometry optimization are a=5.97 Å, b= 6.04 Å and c= 6.99Å. Meanwhile, for the defective 3x2x1 supercell, the lattice constants after geometry optimization are a=6.28Å, b= 6.27 Å and c= 6.22Å. Those results indicate that the lattice structure expands along both the *a*- and *b* axis but shrinks along the c-axis with the presence of vacancy defects. Notably, our DFT calculations are unable to estimate V-V effective interactions, which may require use of the density functional tight binding model. Nevertheless, our DFT calculation results show that total energy of the system is high when the vacancies are distributed along a-axis. This means that the binding interaction of vacancies with respect to atoms nearby is also strong. Thus, the relaxation induced extension along a-axis become negligible. With the consideration of Poisson effect, out-of-plane thickness (or c axis) becomes smaller, and relaxed lattice is close to the original lattice before relaxation. In contrast, vacancies with more distribution along b-axis would lead to a lower energy. The corresponding extension along a- and b-axis becomes longer and compression along c-axis become more noticeable. As a result, the relaxed lattice drastically deviates from the original one and lattice constants along three axes become comparable, giving rise to quasi cubic lattice.

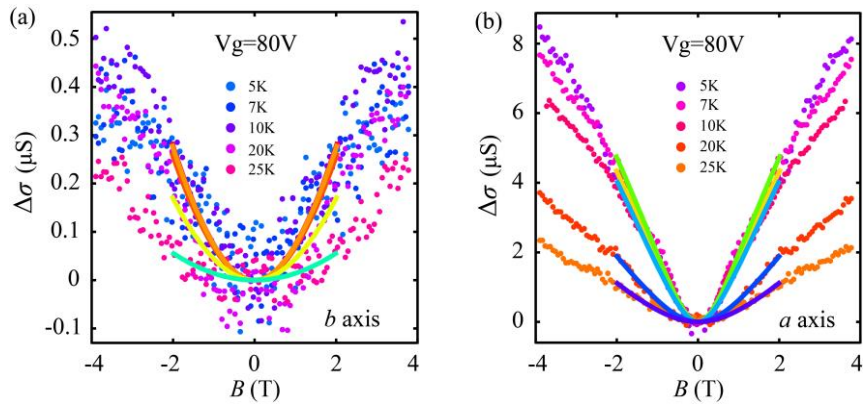

**Fig. S8. The temperature-dependent weak localization effect for *a*-axis and *b*-axis.**